Silver staining of proteins in polyacrylamide gels: a general overview

Thierry Rabilloud*, Laurent Vuillard[+], Claudine Gilly and Jean Jacques Lawrence

CEA - Laboratoire de Biologie Moléculaire du Cycle Cellulaire,
INSERM U309
DBMS / BMCC
CEN-G, 17 rue des martyrs
F-38054 GRENOBLE CEDEX 9 FRANCE

(Running title): overview of silver staining methods for proteins

[+]: Institut Laue-Langevin , BP156, 38042 Grenoble Cedex 09, France.
*: to whom correspondence should be addressed

Correspondence :
Thierry Rabilloud DBMS/BMCC
CEN-G, 17 rue des martyrs,
F-38054 GRENOBLE CEDEX 9
Tel (33)-76-88-32-12
Fax (33)-76-88-51-00

ABSTRACT

On the basis of the physico-chemical principles underlying silver staining of proteins, which are recalled in this paper, several methods of silver staining of proteins after SDS electrophoresis in polyacrylamide gels or isoelectric focusing were tested. The most valuable protocols are presented in this report, including standard methods for unsupported gels and new methods devised for thin (0.5mm) supported gels for SDS electrophoresis or isoelectric focusing and for staining of small peptides. Generally speaking, the most rapid methods were found to be less sensitive and less reproducible than more time-consuming ones. Among the long methods, those using silver-diammine complex gave the most uniform sensitivity. They require however special home-made gels, and cannot be applied to several electrophoretic systems (e.g. systems using Tricine or Bicine as the trailing ion, or isoelectric focusing in immobilized pH gradients). For these reasons, protocols based on silver nitrate are of a more general use and might be favored. Future trends for silver staining will also be discussed.

Abbreviations:

AcOH: acetic acid; AcOK: potassium acetate; AMPSO:3-[(1,1 dimethyl-2hydroxyethyl) amino] 2-hydroxy 1 propanesulfonic acid; APS: ammonium persulfate; BES: N,N-Bis (hydroxyethyl)amino ethane sulfonic acid; Bis: N-N' methylene bis acrylamide; Bicine: N,N-Bis (hydroxyethyl)glycine; %C: ratio (in percent) of crosslinker to total monomers; CHAPS: 3[ (3-cholamidopropyl) dimethylammonio] propane sulfonate; DTT: dithiothreitol; EtOH: ethanol; HEPES: 4-(2-hydroxyethyl) 1 piperazine ethane sulfonic acid; IAM: iodoacetamide; IEF: Isoelectric focusing; IPG: IEF in immobilized pH gradients; MOPS: morpholino propane sulfonic acid; NDS: 2,7 naphtalene disulfonic acid; PAGE: polyacrylamide gel electrophoresis; PDA: piperazine diacrylamide; PBS: phosphate buffered saline; PMSF: phenyl methyl sulfonyl fluoride; POPOP: 1,4 bis (Phenyloxazolyl) benzene; SDS: sodium dodecyl sulfate; STS: sodium thiosulfate pentahydrate; %T: grams of monomers (including crosslinker) per 100 ml of gel; TES: tris(hydroxymethyl)aminoethanesulfonic acid; Tricine: N-tris(hydroxymethyl) glycine; Tris: Tris(hydroxymethyl)aminomethane

INTRODUCTION

The introduction of silver staining of proteins in polyacrylamide gels in 1979 (Switzer et al. 1979) has been a breakthrough in the field of protein detection, raising the sensitivity from the microgram range obtained with dyes such as Coomassie blue to the nanogram range. This breakthrough as well as the numerous problems encountered with this first protocol (cost, background, non specific surface staining) led to extensive research and to the description of various silver staining methods (reviewed in Rabilloud, 1990). To summarize these data, we can consider that every staining protocol is divided into five main phases.

In the first phase, the gel is fixed to eliminate the interfering substances otherwise giving a high background and therefore poor contrast (e.g. SDS, ampholytes, amino acids, Tris).

In the second phase, the gels is sensitized, i.e. treated with an agent which will eventually increase the sensitivity and/or the contrast.

In the third phase, the gel is impregnated with the silvering agent. Two main families of silver staining methods can be distinguished at this step, those using silver nitrate solutions as the silvering agent (acidic methods) and those using a basic silver-ammonia or less frequently a silver-amine complex (basic methods).

In the fourth phase, the image is developed, generally by a dilute formaldehyde solution. In the last phase, the reaction is stopped to prevent overdevelopment and the gel is prepared for storage. The basic physico-chemical principles underlying these methods will be recalled in the next section. Based on these principles, selected silver staining protocols will be described and some results shown. The future trends for silver staining will be finally discussed

2. SILVER STAINING PRINCIPLES

The silver staining principles have been detailed in a previous paper (Rabilloud, 1990). They will be just briefly summarized here

2.1 Silver image formation

Two main phenomena underline the formation of a silver image. The first one is the key of the photographic processes and is the autocatalytic reduction of silver. This means that the reduction of silver ion into metallic silver is greatly accelerated when some metallic silver is already present (latent image). This phenomenon can be extended to some other chemicals such as other rare metals (gold, palladium platinum) and some heavy metals sulfides (silver, lead, mercury sulfides for example). When applied to image formation in silver staining, this means that the image will appear at the places where the reduction begins, since image growth will be much faster there than anywhere else. In the case of silver staining of macromolecules, this is unfortunately in contradiction wioth the second principle, which is that the tighter the complex between silver ion and a molecule, the less reactive the silver ion. The simple application of these two principles in a silver staining protocol is outlined in figure 1 an results in negative staining, as shown by Merril's group (Merril and Goldman 1984).

In order to obtain a positive stain, one must use the complexing power of proteins for silver ion. This is easily done by using a very weak developer based on formaldehyde at alkaline pH, and by rinsing the gel after the silver impregnation and before development. The result is schematized on figure 2. During the rinse, the silver ion or complex will diffuse out of the gel much faster from the places where the complexation is weak (background) than from the places where the complexation is strong (protein). The key of silver staining lies in the fact that there is a point (developing point) where the resulting excess of silver ion at the protein level will increase the speed of reduction to a point where it will overcome the slowing effect due to complexation. The reduction will therefore begin at the location of proteins and therefore ensure a positive image through the autocatalytic phenomenon desribned above. The post silver rinse is a critical step which must be carefully timed. Too much rinse will eventually decrease the silver ion concentration and lead to a reduced sensitivity, whereas insufficient rinsing will give a high background. On an other hand, the speed of reduction is also very critical. It has been shown that only certain aldehydes at a defined alkaline pH (11 to 12) can produce a positive image (Heukeshoven and Dernick 1985). To achieve such conditions, dilute solutions of formaldehyde in sodium or potassium carbonate are used with silver nitrate-treated gels (Merril and Goldman 1984). With silver-ammonia-treated gels, solutions of formaldehyde containing very small amount of citric acid are used (Oakley et al 1980). This acidification is necessary because of the very high pH prevailing in the silver -ammonia bath (ca. 13) leadind to a too high reducing activity of formaldehyde. This results in poor contrast and high background. Because of this acidic development bath, the absolute reduction speed within the gel is degressive since the progresive diffusion of the acid into the gel will decrease the activity of the developer. However, the autocatalytic activity ensures sufficient image formation under these conditions.

This diffusion-driven staining mechanism also explains why silver staining is so difficult in agarose gels or in immobilized pH gradients gels. In these cases, the gel itself significantly binds silver ion, yielding a poor contrast between the protein zones and the background, as well as a poor sensitivity.

Another important consequence of this diffusion-driven mechanism is the variety of colors of the developed silver. It has been shown (Merril et al 1988) that this color depends on the size of the silver grain. The bigger the silver grain, the darker the color. This size of the silver grain is dictated by the speed of reduction, a faster reduction leading to a bigger grain. This fact, as well as the mechanism outlined in figure 2, explain the color effects (chromatism) frequently encountered in silver staining.

When proteins are present at a very high concentration (e.g. albumin in plasma) the center of the gaussian electrophoretic band or spot will contain a very high amount of protein, able to achieve a tight binding of silver ions. Oppositely, the edges of the zone will contain much less protein and then will bindi a lesser amount of silver ion. In this diffusion-driven mechanism, the edges of the zone will reach the developing point earlier than the center of the zone, where the complexation of the silver is so high that it prevents silver reduction. Due to the autocatalytic process, this difference between the center and the edges will be amplified and lead to a fast reduction and therefore to a dark color on the edges, whereas limited reduction will take place at the center, giving a light color. In extreme cases, the complexing power of the protein for silver ion is so high that the developing point cannot be reached (Yüksel and Gracy 1985), leading once again to negative staining. The chromatism effects can therefore be explained by various abilities of different proteins to release the silver ion during the development step, leading to different reduction speeds and therefore to different colors.

It must be stressed that such simple silver staining protocols are not optimal. For silver nitrate staining, the sensitivity is very low, while acceptable results are obtained with silver-ammonia staining (Wray et al 1981), although many proteins are weakly stained and some are not stained at all. To alleviate these problems, the gel is very often treated with various reagents prior to silver impregnation in order to increase both sensitivity and contrast and to obtain a more uniform staining of different proteins. This sensitization process will be described below.

2.2 Sensitization

Owing to the basic principles described above, two main methods can be envisioned to sensitize the gel.

In the first method, outlined in figure 3, the gel is treated with a reagent which is adsorbed by proteins and which then binds silver ion or silver complexes. The gel is then rinsed in order to remove the reagent from the background. The gel is then impregnated with the silvering agent. This treatment provides extra binding sites for the silvering agent, and there is more silver ion adsorbed to the protein. Upon development, this extra silver ion will be reduced, which will increase the sensitivity. The effect is generally magnified by the fact that these reagents bind silver less avidly than proteins. This implies in turn that silver ion bound to the sensitization reagent will be easily reduced. The reduction will therefore start earlier and more efficiently at the protein level, leading to increased sensitivity and contrast.

Such reagents are listed in Table 1. It must be stressed that silver-ammonia methods can use several different reagents whereas silver nitrate methods can use only SDS and the agents containing nucleophilic groups such as Coomassie blue.

The second method to sensitize the gels (which can be used in combination with the former) is schematized on figure 5. Here, the gel is treated with a reagent exhibiting some affinity for the proteins and able to form either metallic silver or silver sulfide. With this sensitization process, the latent image will be formed during the silvering step, so that silver reduction will be very fast at the protein level, ensuring good sensitivity and contrast.. It must be stressed that the sulfiding agents (listed in table1) are generally only used in silver nitrate methods. In the silver-ammonia methods, the concentration of free silver ion is very low, thus reducing the formation of silver sulfide. As opposed to the sulfiding agents, the reducing agents (both listed in table 1) can be used with both types of methods. It must however be stressed that the aldehydes can reduce silver only at an alkaline pH which makes them efficient as sensitizers for in silver-ammonia methods only.

2.3 An other way to increase sensitivity: the use of thiosulfate

Despite the use of various sensitizers in silver staining, the contrast and the sensitivity of silver staining remain mostly inadequate because of a strong tendency to background increase, imposing premature termination of the development (i.e. before optimal sensitivity can be reached). This is due to the fact that minute silver deposition at the background level will give strong background because of the autocatalysis. The remedy to that problem is to use very low amounts of thiosulfate during development. Thiosulfate at neutral and basic pH is able to give very strong complexes with silver, since it is able to dissolve silver halides. Such strong complexes are therefore not prone to reduction in the weak conditions used for silver staining. By adding micromolar amounts of thiosulfate in the developing bath of a silver nitrate-based method, it has been shown ( Blum et al 1987) that a dramatic reduction of the background is obtained, probably because the thiosulfate prevents any spurious silver or insoluble silver salts to deposit, thereby cleaning the background. The same effect can be obtained in silver-ammonia-stained gels by adding thiosulfate to the gel at the polymerization stage (Hochstrasser and Merril 1988). In silver ammonia-stained gels, it is not possible to add thiosulfate into the developing bath because the penetration of the thiosulfate will sequester the silver present and prevent image formation. This is due to the fact that the acidic developing bath gives a lower reduction speed than the alkaline bath used with silver nitrate. As opposed to citric acid developers, carbonate-based developers favor image formation over its prevention by thiosulfate.

In conclusions, the features required for a modern and efficient silver stain are:

-the presence of a sensitizing step, using preferably both types of sensitization

-the use of thiosulfate in the development process

On these bases, numerous protocols have been tested in various electrophoretic techniques (Rabilloud 1992; Rabilloud et al 1992). The most performant protocols in each class will be detailed in the following sections.

3. MATERIALS AND METHODS

3.1 Sample preparation.

3.1.1 SDS-PAGE

Bio-Rad molecular weight markers (high and low ranges) were diluted into sample buffer (Bis-Tris 0.24M pH 6.5; SDS 2%; glycerol 20%, DTT 50mM) and boiled for 5 minutes. When alkylation of the sample was performed, 1/20th volume of iodoacetamide (IAM) stock solution (4M in methyl formamide (MF), made freshly by adding 1g IAM to 1 ml MF )was added and the mixture left at room temperature for 30-120 minutes before loading onto the gel. Serial dilutions were performed after alkylation in sample buffer minus DTT.

3.1.2 Isoelectric focusing

3.1.3 Two-dimensional electrophoresis

Plasmocyte cells (P3-X63-Ag8), grown in RPMI 1640 medium supplemented with 10% FCS and 50μM mercaptoethanol, were collected and washed twice in PBS. The cell pellet ($10^8$ cells) was resuspended in 40μl of solution A [CHAPS 20% (w/v), thioglycerol 50% (v/v)]. 360μl of solution B [urea 10M, Tris-phosphate pH 7 50mM, PMSF 1mM] made fresh and kept until use at 37°C were added while vortexing. After extraction for 1 hour at room temperature with occasional vortexing, the viscous mixture was centrifuged at 200000 g for 1 hour to pellet the DNA. 10 μl Pharmalytes 3.5-10 were added to the supernatant, and the protein concentration was determined by a modified Coomassie blue assay (Ramagli and Rodriguez 1985).

3.2. Gel electrophoresis

3.2.1. SDS PAGE

SDS PAGE was generally performed in a discontinuous gel system (gel dimension 190 x 160 x 1.5 mm) with the following buffer compositions:

Resolving gel: Tris-HCl 0.33M pH 8.4

Stacking gel: BisTris-HCl 0.24M pH 6.5

Electrode buffer: 100mM Tris, 100mM taurine, 0.1% SDS

Some tests were also carried out with different electrophoresis systems using different trailing ions such as glycine (Johnson 1982), Tricine or borate (Patton et al 1991), Bicine (Wiltfang et al 1991), and homemade systems using Tris-HCl buffers at a pH selected to give a relative mobility of -0.2 to -0.23 relative to sodium ion with the following trailing ions: Asparagine, Mops, Hepes, Ampso, Tes, Bes.

The stacking gels were acrylamide-Bis gels (4.5%T, 2.6% C), but two types of resolving gels were used depending on the silver staining method to be used.

For gels to be stained with silver nitrate methods, standard acrylamide-Bis gels (10%T, 2.6% C) were used. The gel mixture was filtrated (0.2μm) before use but not degassed. Polymerization was initiated by adding 500μl APS10% and 25μl Temed per 60ml gel mixture. For gels to be stained with silver diammine methods, acrylamide-PDA(Bio-Rad) gels were used (10%T, 2.6%C), with thiosulfate added to the initiator system (Hochstrasser et al 1988b;.Hochstrasser and Merril 1988) After filtration of the gel mixture, polymerization was initiated by the sequential addition of 500μl of sodium thiosulfate pentahydrate (STS)10%, 50μl Temed and 500μl APS 10% per 60ml of gel. These acrylamide-PDA gels were also tested with silver nitrate methods but were not further used, because no improvement was observed in this case over standard Bis crosslinked gels.

In some cases, the gels were photopolymerized (Lyubimova et al. (1993)) was used. In this case, the initiator sytem was (final concentrations in gel) 25μM diphenyliodonium chloride, 500μM sodium toluene sulfinate and 20μM methylene blue. Polymerization was initiated by irradiating the gel with visible light (500W halogen lamp at a 40 cm distance). In some cases, sodium thiosulfate (1mM final concentration) was added to the gel mixture prior to polymerization.

Separations were carried out in a Protean II cell (Bio-Rad) thermostated at 4°C and operated at a constant current of 40mA/gel, which resulted in 5 to 6 hours migrations.

3.2.2. Peptide separations

For the separations of small peptides, an electrophoretic system based on the Tris-HCl-Tricine system (Schägger and Von Jagow 1987). The resolving gel had the following composition: Acrylamide-Bis 16.5%T, 6%C in 1M Tris-HCL pH 8.45 and 11% glycerol. It was overlaid with a spacer gel (10%T, 3%C in the same buffer without glycerol). A stacking gel (4%T, 3%C in 0.75M Tris-HCL pH 8.45) was cast on the top.

The anode buffer was Tris-HCl 0.2M pH 8.9 and the cathode buffer was Tris 0.1M, Tricine 0.1M SDS 0.1%

For a 12cm separating gel overlaid with a 2cm spacer gel and a 1.5 cm staking gel, the running conditions were: 2 hours at 30V, then 18hours at 120V and 4 hours at 320V

3.2.3. Isoelectric focusing

Starting from a commercial pH 4 to 7 immobilized pH gradient plate (Pharmacia), individual 3 mm wide strips were cut and reswollen overnight at room temperature in 0.4% carrier ampholytes (Pharmacia 3-10). The sample (5-20µl in the rehydrating medium) was applied on a paper applicator (Pharmacia) at the cathodic side of the gel. Running conditions were the following: 100V, 1hr; 300V 1 hr; 500V 1 hr; 800V overnight.

3.2.4. Two-dimensional electrophoresis

Two-dimensional electrophoresis was performed essentially as described by O'Farrell (1975) with some modifications (Hochstrasser et al 1988a), using 1.5 mm capillary tubes in the first dimension for the IEF (carrier ampholytes generated gradients). IEF was carried out for 16000 Vh. The extruded rod gel was held on the top of the second dimension gel without any agarose (Hochstrasser et al 1988a). The second dimension gel was a continuous 10% T, 2.7%C gel without stacking (Hochstrasser et al 1988a), including PDA and thiosulfate when silver diammine staining was to be carried out. Running conditions as for standard SDS gels (see 3.2.1)

3.3. Silver staining

The silver staining methods tested were classified into four types:
-fast methods (completed in less than 5 hours) using silver nitrate
-long methods using silver nitrate
-methods using silver diammine
-methods devised for thin supported gels (0.5mm) either for SDS PAGE or IEF

3.3.1. General features

All the chemicals used were analytical grade and were generally purchased from Fluka (unless otherwise indicated), except ethanol which was technical grade absolute ethanol and HEPES hemisodium salt coming from Sigma. Deionized water with a resistivity higher than 10MΩ/cm was used. Gels were stained in plastic containers with up to 4 gels per container unless otherwise indicated. The solution to gel ratio was 250ml/gel (except for silver diammine (100ml/gel)), with a minimum of 500 ml per container (silver diammine 300ml).

3.3.2. Fast methods for unsupported SDS gels

Previous tests including the protocols devised by Merrill et al (1981), Hempelmann et al (1984), Morissey (1981), Doucet and Trifaro (1988), Blum et al (1987) and Rabilloud et al (1988), have been described elsewhere (Rabilloud 1992). It appeared that the first three protocols were plagued with either a low sensitivity and/or a high background (even with the addition of thiosulfate as a background preventing agent ). Among the latter protocols, the test showed that the best methods were those of Blum et al.(1987) and Rabilloud et al.(1988) the latter being slightly superior to the former. The protocols for these methods are given in Table 2. The variant in table 2 column C is designed for the staining of small peptides in a Tricine-based electrophoretic system avoiding any fixation in acid-alcohol mixtures which could lead to leakage of the peptides from the gel.

3.3.3. Long (>5 hours) silver nitrate methods.

When compared to the fast methods, these protocols often use an overnight fixation step and extensive washes to remove the excess sensitizing agent. Although its sensitivity is claimed to be very high (Ochs et al 1981) the method of Sammons et al (1981) was not tested since it consistently gives a deep yellow background which is often not desirable. The basic method in this family is therefore the one described by Heukeshoven and Dernick (1986). We described recently a variant which proved more sensitive and less prone to background development. This method is detailed in Table 3 col A.

3.3.4. Silver diammine methods.

These methods are generally based on the work of Oakley et al (1980), but were recently dramatically improved (Hochstrasser et al 1988b;.Hochstrasser and Merril 1988). The improved method of Hochstrasser et al.(1988b) was used as a basis. We however found that the addition of NDS in the sensitization process gave improved sensitivity, especially for rather acidic proteins such as soybean trypsin inhibitor which are otherwise poorly detected. This improved staining protocol is described in Table 3 col B.

3.3.5. Protocols for thin supported gels

The basic protocol for staining thin supported gels is the one devised by Heukeshoven and Dernick (1988). However, it was shown previously that this protocol could be greatly improved by adding thiosulfate to the developing agent (Rabilloud et al. 1992). The latter protocol (listed in Table 4 col. 1) is quite adequate for thin SDS gels. For gels which do not contain SDS, further increase in sensitivity and resolution is afforded by replacing the acid-alcohol fixation by a TCA fixation (Table 4 col.2).

4. RESULTS

4.1. Sensitivity tests on MW markers in SDS PAGE

The selected protocols, generally issued from a previous comparative study (Rabilloud 1992) were tested with molecular weight markers in different systems. Owing to the recent development of electrophoretic systems which do not use glycine as the trailing ion (Patton et al. 1991; Schägger and Von Jagow 1987; Wiltfang et al 1991) we tested these various protocols on such sytems. It appeared that the silver nitrate methods (Tables 2 and 3A) were not sensitive to the nature of the trailing ion, provided that correct fixation (at least 3 x 30 minutes) was carried out. In all the systems tested, results were quite similar to the one shown in figure 5A, with detection limits in the lower (2-10) nanogram range. A result obtained with small peptides is shown in figure 6, showing that the protocol modified for does not give rise to any increased background.

Oppositely, the silver-ammonia methods proved to be very sensitive to the trailing ion. High background preventing the use of such methods was obtained with the following trailing ions: MOPS, BES, TES, HEPES, Tricine and Bicine

Correct results were obtained with simple amino acids (glycine, taurine, asparagine) with borate and with AMPSO, the latter two ions showing however decreased electrophoretic resolution. A typical result is shown on figure 5B. Here again, the detection limits are in the low nanogram range.

Concerning gels polymerized with the photoinitiating system of Lyubimova et al (1993), results equivalent to those obtained with persulfate-polymerized gels were obtained with silver nitrate methods. This was not the case however with silver-ammonia methods. In the case on thiosulfate-free gels, a huge surface staining developed quite early, preventing the use of the methods. When 1mM (half the concentration used in Hochstrasser and Merril 1988) was included in the gels. The sensitivity dropped dramatically (data not shown). We interpret this phenomenon by a much greater grafting of the thiosulfate in the gel with the photoinitiating system, leading to excessive complexation of the silver ion by the gel and thus reduced sensitivity.

4.2. Tests in isoelectric focusing gels

Typical results obtained for staining of IEF-IPG gels are shown on figure 7. When compared to the results obtained with SDS gels, it can be seen that the stains are much less sensitive (detection limits ca. 50ng instead of <10 ng) and quite variable from one protein to another. This is related to the fact that staining of native proteins is known to be mauch more, variable and less sensitive than staining of SDS-denatured proteins (Marshall 1984)

4.3. Tests on complex mixtures

In order to assess the homogeneity of the stains (i.e. the protein to protein variation) and their proneness to artefacts such as hollow or negative staining, further testing was performed using two dimensional electrophoresis of acidic, neutral (O'Farrell 1975) and basic proteins (Rabilloud in press). Typical results are shown on figures 8 and 9. For acidic and neutral proteins (figure 8), both methods give comparable sensitivity. It can be observed that the intensity of a same spot is not conserved from one method to another. However, we could not find a spot detected by a method and not by the other one. This is however not true with basic proteins, as shown in figure 9. Here, the silver-ammonia methods proved much more sensitive than the silver nitrate methods. Silver ammonia methods also gave darker spots and were less prone to "hollow" staining. The tendency towards hollow staining seems to be strongly dependent on silver concentration, decrearing from high with methods using 0.1% silver nitrate,to moderate for methods using 0.2% silver nitrate  and very low for silver diammine methods using silver concentration of 0.4 to 0.8%, provided that the development period is long enough. Indeed, hollow spots were observed at the early stages of image development, but were reversed into regular dark spots as the development proceeded. It was therefore essential to use long (at least 10 minutes) development times to obtain consistent results and the developing solutions were  modified for this purpose.

DISCUSSION

The main problems associated with silver staining at the early stages of the method were
- the high and erratic background (thereby decreasing the sensitivity),
-the extreme protein to protein variability in staining and the wide range of colors obtained for various proteins, leading in some cases to "hollow staining"with a transparent or weakly colored center of the protein zone and a more deeply colored periphery.
These drawbacks impaired the use of silver staining limited as a routine procedure, as well as prevented any real quantitative use of the data obtained. These difficulties in the use of silver staining were exemplified by the great number of different protocols which appeared in the litterature in the early 80's (more than 60), suggesting that no protocol gave very adequate results.
.
A landmark in silver staining was to introduce the use of thiosulfate to sequester silver ion weakly bound to the matrix, thus preventing its spurious reduction at the development stage and thereby dramatically reducing the background. As pointed out by Blum et al. (1987) and further examplified by Rabilloud et al. (1988), thiosulfate has allowed to use very powerful sensitizing compounds which would have been impossible otherwince, as this would have led to excessive background.

A first protocol with thiosulfate as background reducer using a silver diammine staining was described by Wiederkehr et al (1985), but the post-silver thiosulfate step had to be very carefully timed to achieve reproducibility. A much more convenient protocol was described by Blum et al (1987) for silver nitrate methods. There, thiosulfate was directly added to the developing solution. Such an approach could not be used for silver diammine methods, since it would lead to a dramatic decrease in sensitivity.

A convenient method for using thiosulfate in silver diammine methods has been recently described by Hochstrasser and Merril (1988), who included thiosulfate in the gel formulation to graft it into the matrix upon polymerization. It must be emphasized, however, that the success of this copolymerization approach is very dependent on the polymerization conditions, since the use of a different initiating system (photopolymerization) gave completely different results, showing that an adaptation of the thiosulfate concentration in the gel is mandatory. Moreover, this approach cannot be used with gels polymerized under acidic conditions, since the thiosulfate decomposes under pH 4, generating sulfide, a polymerization inhibitor.

The most surprising fact observed in the course of this study is the wide divergence existing between silver nitrate and silver-ammonia methods in their behavior in different electrophoretic systems. The first important point is the requirement for a different crosslinker, namely Piperazine Diacrylamide (PDA) (Hochstrasser et al 1988b). This can be easily explained when taking into account the very basic conditions prevailing in the silver ammonia bath (pH 13). Under such conditions, there is an alkaline hydrolysis of the matrix. While acrylamide hydrolyzes into acrylic acid and ammonia, having no effect on the staining process, the hydrolysis of Bis will produce formaldehyde (see figure 10) thus reducing silver and producing a latent image throughout the gel turning into a high background upon development. This noxious hydrolysis does not arise with PDA, which is more resistant than Bis to alkaline hydrolysis because it is a tertiary amide. Moreover, the hydrolysis products of PDA (piperazine and acrylic acid) do not induce background formation in silver staining.

Another problems derives from the fact that various trailing ions cannot be used in silver ammonia methods. Several hypotheses can be taken into account.

i) this is due to a side reaction between the trailing ion remaining in the gel and the glutaraldehyde used for fixation. This hypothesis can be ruled out easily because some trailing ions such as Bicine and BES bear tertiary amine moieties (which are much less reactive toward glutaradehyde than primary amines (glycine, taurine) are) and still give rise to high background. This is easily shown when direct fixation in glutaraldehyde after electrophoresis is performed. In gels containing primary amino acids (e.g. glycine) a deep brown color develops immediately because of the reaction of glutaraldehyde with glycine. This color will persist in the gel during the washing process and will be converted into high background upon silver staining. This does not happen with Tricine (figure 6), so that the hypothesis of a glutaraldehyde effect can be ruled out.

ii) The second hypothesis is based on the assuption that complex trailing ions (e.g. Tricine) are less efficiently washed out from the gel than simpler ones (such as glycine). Owing to the high concentrations of the trailing ions present in the gels at the end of the electrophoresis run (over 0.1M) some is kept in the gel up to the silver impregnation. In the case of silver nitrate methods, it must be assumed that the complex between these compounds and silver does not interfere with subsequent staining or that any negative effect is masked by the use of thiosulfate in the developer. In the case of silver-ammonia methods, an orange-brown color develops already in the silver-ammonia bath. This indicates that, under these very alkaline conditions, the trailing ion is able at least to form an unstable complex with silver. This hypothesis is favored by the fact that many of these complex trailing ions are known to exhibit high affinities for heavy metal ions. For example the Ka of Tricine with copper ion is over $10^7$, implying a strong binding, which will be even increased at very high pH, were the ionization of the OH group begins to play an synergistic role with the amino group. In the case of hydroxylated trailing ions (Tricine, Bicine), another phenomenon can take place. at very alkaline pH, the alcohol can be oxidized by the dissolved oxygen to give the aldehyde and the acid. This process can also be driven by the presence of silver ion and induce silver ion reduction and thus background formation.

Another major difference between the two types of staining concerns the detection of prteins with very acidic or very basic pIs. in the case of acidic proteins (e.g. soybean trypsin inhibitor) the silver nitrate stain are superior to the ammoniacal stain. This is probably linked with the rather low lysine content of these proteins, since a positive correlation between the lysine content and the staining with ammoniacal methods has been described (Dion and Pomenti 1983). The opposite is true for basic proteins, where the silver nitrate methods are quite not optimal. This can be explained by the fact that at the acidic pH of the silver nitrate solution (ca. 6) the basic proteins behave as cations, which induces an electrostatic repulsion between the protein and the silver ion, thereby decreasing the extent of silver bound and thus the staining intensity. This hypothesis is supported by the fact that the addition of HEPES buffer in the silver bath enhances the detection of basic protein with silver nitrate methods, although the sensitivity remains lower than the one of silver-ammonia methods (result not shown). It must be stressed at this point that pH must be carefully controlled, since spurious reduction of the silver nitrate by the formaldehyde of the bath begins at pH 7.8. HEPES hemisodium salt was therefore used to avoid any problem coming from inaccurate pH adjustment.

In the case of isoelectric focusing gels with immobilized pH gradients, a further problem comes from the grafting of the buffering groups within the gel. This is especially a problem with basic compounds which bear tertiary amine functions which bind silver quite comparably to proteins. This induces a gradient in background increasing from the acidic to the basic part, which is especially pronounced with ammoniacal silver methods where a silver mirror is frequently obtained (Rabilloud et al 1992). We therefore focused our effort on silver nitrate methods. Based on a previously tested protocol (Rabilloud et al 1992), we tested several compounds to try to increase the sensitivity and the contrast. These compounds included thiourea, phenylthiourea, dithionite , thiosulfate and tetrathionate, already tested for SDS gels (Rabilloud et al 1988). Of these tetrathionate proved superior to thiosulfate, which was itself clearly superior to the other compounds. This superiority of tetrathionate over thiosulfate, also observed for SDS gels (Rabilloud 1992), is probably linked to the difference in the stability of the two compounds. It is well known that thiosulfate is very unstable at moderately acid pH, with strong decomposition at pH 4 and below with generation of sulfide which is a very powerful background promoting agent, since the hydrogen sulfide produced at low pH has probably a very weak (if any) affinity for proteins. As the fixation process generally uses very acidic pH, the thiosulfate sensitization must be carried out at a controlled pH (Heukeshoven and Dernick 1988). It is however not sure that the buffering effect coming from the diffusion of the buffer into the gel will always be fast enough to increase the pH everywhere in the gel before any thiosulfate decomposition occurs. This could explain why some erratic results with a high background are sometimes obtained with thiosulfate. Tetrathionate, on the other hand, is much more stable to the acid, so that this spurious decomposition does not occur. Moreover, tetrathionate is a much bigger anion that thiosulfate. If we assume that proteins behave to some extent as weak ion exchange materials, this combination of large size with multivalency (-2 both for thiosulfate and tetrathionate) is very favorable for binding. This more efficient binding of tetrathionate would also explain the better results obtained with this ion.

We would like to conclude this discussion in adressing the future developments of silver staining. The principal aim would be to break the "nanogram barrier", since, at least in SDS gels, all the methods tested to date reproducibly give a detection limit of a few nanograms. In order to increase the sensitivity, we have tried to increase the latent image formation by coupling covalently the proteins (before or after electrophoresis) to compounds able to give silver sulfide (e.g. isothiocyanates) or to strongly reducing compounds (e.g. polyhydroxy aromatic compounds). The latter class was chosen on the basis of a work showing a much faster image formation for proteins bearing hyperhydroxylated tyrosine residues (Wells and Cordingley 1991). The results were however rather disappointing. The image formation was indeed very fast, but the final results were not superior to those obtained with the methods described here. This suggested that the current limiting factor in silver staining is not the latent image formation, but the extent of the silver binding. In our model of silver staining, this would imply that the formation of the image uses mainly (and maybe only) the silver bound to the proteins, and not the silver present in the remaining of the gel. We therefore tried to increase the number of binding sites on the proteins either by coupling the proteins with reactive dyes bearing amino , sulfonic and /or heterocyclic groups prior to electrophoresis (Bosshard and Datyner 1977) or by impregnating the gels after electrophoresis with various sulfonated dyes such as Coomassie blue (De Moreno et al. 1985), copper phtalocyanine tetrasulfonate (Bickar and Reid 1992) or other acidic dyes. No increase in sensitivity was observed for the covalent coupling, whereas some increase in sensitivity was observed for the post electrophoretic treatments, at the expense of a great increase in the background precluding the routine use of such techniques. The latter effect can be explained by the fact that such dyes effectively bind silver, but with such an affinity that ecven minute amounts of dye remaining in the gel will give background. The problem turns therefore to the search of conditions where the concentration of the dye within the gel will be very low (which implies more than numerous washes) while some dye will still be bound on the proteins to increase the binding. This requires dyes with a very high affinity for proteins and a very high solubility. Coomassie dyes are therefore good candidates, but their affinity proved too low to meet the requirements to break the nanogram barrier. On the other hand, preelectrophoretic coupling of reactive dyes gave no increase in sensitivity. This is probably due to the fact that the sites where these dyes react (Lys, Cys, Tyr) are already binding sites for silver in the proteins. This means that the increase in silver binding is only marginal with such a process, leading to no increase in sensitivity.

This means that silver staining may have reached a plateau in its sensitivity, which would be broken only if different mechanisms for staining can be found, since the affinity of proteins for silver is now fully used.

Table 1: Main sensitizing agents

| Mode of action and nature of the reagent | used in silver nitrate staining | used in silver ammonia staining |
|---|---|---|
| **INCREASED SILVER BINDING** | | |
| Coomassie Blue R or G 250 | + | + |
| Naphtalene disulfonic acid | - | + |
| SDS | + | + |
| Sulfosalicylic acid | + | + |
| Di sulfo POPOP | + | - |
| **SILVER SULFIDE FORMATION** | | |
| Sodium thiosufate | + | - |
| Potassium tetrathionate | + | - |
| Farmer's reducer (thiosulfate + ferricyanide) | + | - |
| (phenyl)thiourea | + | - |
| **SILVER ION REDUCTION** | | |
| Sodium phosphite and hypophosphite | + | - |
| borohydrides | + | - |
| low valency copper or iron salts | + | - |
| formaldehyde | - | + |
| glutaraldehyde | - | + |
| **UNCLEAR MECHANISM** | | |
| DTT | + | - |

Sodium dithionite + -

Table 2: Rapid silver nitrate protocols for staining unsupported SDS gels

| PROTOCOL | BLUM et al (1987) | RABILLOUD et al (1988) | small peptides |
|---|---|---|---|
| FIXATION | AcOH 10% EtOH 30% **3x30'** | AcOH 10% EtOH 30% **3x30'** | Glutaraldehyde 5% **1hr** |
| RINSE | EtOH 20% **10'** then water **10'** | EtOH 20% **10'** then water **10'** | water **6x20'** at least* |
| SENSITIZATION | STS 0.2 g/l **1'** | dithionite 0.3 g/l **1'** | dithionite 0.3 g/l **1'** |
| RINSE | water **3x20''** | water **2x1'** | water **2x1'** |
| | SUBSEQUENT STEPS COMMON TO THE 3 METHODS | | |
| SILVER | silver nitrate 0.2%; 37%HCHO 0.7ml/l : **30-60'** | | |
| RINSE | water **5-10''** | | |
| DEVELOPMENT | Potassium carbonate 3%, 37%HCHO 0.25 ml/l, STS 10mg/l: **5-15'** | | |
| STOP | Tris 50g/l, AcOH 25 ml/l **30'** | | |
| STORE | water, several rinses | | |

Table 3: long methods for staining unsupported SDS gels

| PROTOCOL | SILVER NITATE | SILVER-AMMONIA |
|---|---|---|
| FIXATION | AcOH 10%, EtOH 30% **3x30'** | AcOH 10%, EtOH 30%: **1hr**, then AcOH 10%, EtOH 30%: **15-18hr**. |
| RINSE | | cold water: **15'** |
| SENSITIZATION 1 | glutaraldehyde 0.5%, AcOK 0.5M, EtOH 20%, KTT 2.5-3g/l **Overnight** | glutaraldehyde 0.5% and AcOK0.5M in cold water **30'** |
| RINSE | water 6x20' | cold water **4x15'** |
| SENSITIZATION 2 | | NDS 0.5 g/l **2x30'** |
| RINSE | | water **4x15'** |
| SILVER | silver nitrate 0.2% 37%HCHO 0.7 ml/l **120'** Hepes pH 7.5 5-10mM | silver nitrate 0.75% (45mM) Ammonia 180mM **30'** NaOH 45mM |
| RINSE | water **5-15''** | water **3x5'** |
| DEVELOPMENT | 37% HCHO 0.25 ml/l potassium carbonate 3% STS 12.5 mg/l **10-15'** | 37% HCHO 1ml/m 50% citric acid 0.2 ml/l **5-10'** |
| STOP | Tris 50g/l; AcOH 25ml/l | AcOH 20ml/l, ethanolamine 5ml/l |

The use of cold water in the silver ammonia protocol has been found to decrease the background (Hochstrasser et al 1988b). this could be due to the fact that cold water prevents the spurious grafting of glutaraldehyde onto the polyacrylamide gel itself (see Rabilloud, 1990)

Table 4: protocols for staining of thin (0.5 mm) supported gels

| PROTOCOL | SDS GELS | IEF-IPG GELS |
|---|---|---|
| FIXATION | AcOH 10%; EtOH 30% **3x30'** | TCA 10% **3x30'** * |
| | SUBSEQUENT STEPS COMMONT TO BOTH METHODS | |
| SENSITIZATION | AcOK 0.5M; EtOH 25%; KTT 3 g/l; glutaraldehyde 0.5% **45-60'** | |
| RINSE | Water **4x 15'** | |
| SILVER | silver nitrate 0.2%, HCHO 0.5 ml/l **30'** | |
| RINSE | water **5-10''** | |
| DEVELOPMENT | Potassium carbonate 3%, 37% HCHO 0.25ml/l, STS 10mg/l **5-10'** | |
| STOP | Tris 50g/l, AcOH 25ml/l **30'** | |
| STORE | water, several rinses | |

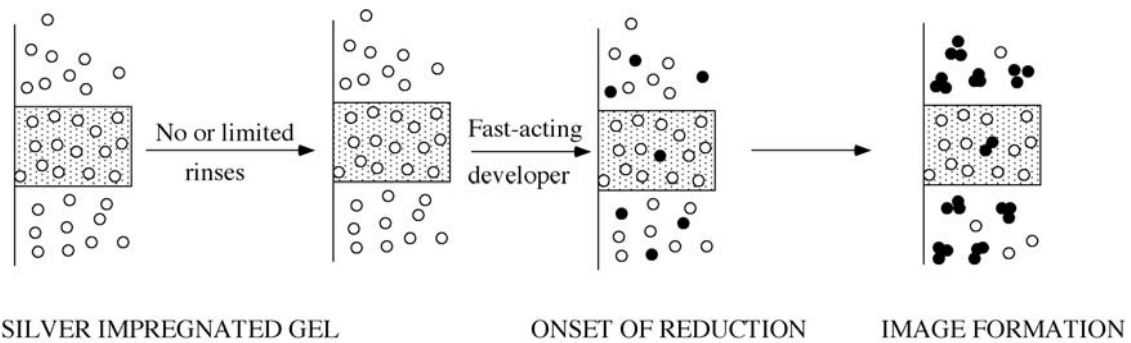

Figure 1:

Schematic representation of a negative silver stain. the dotted zone represents the protein in the gel, the open circles the silver ion, the dark circles metallic silver.

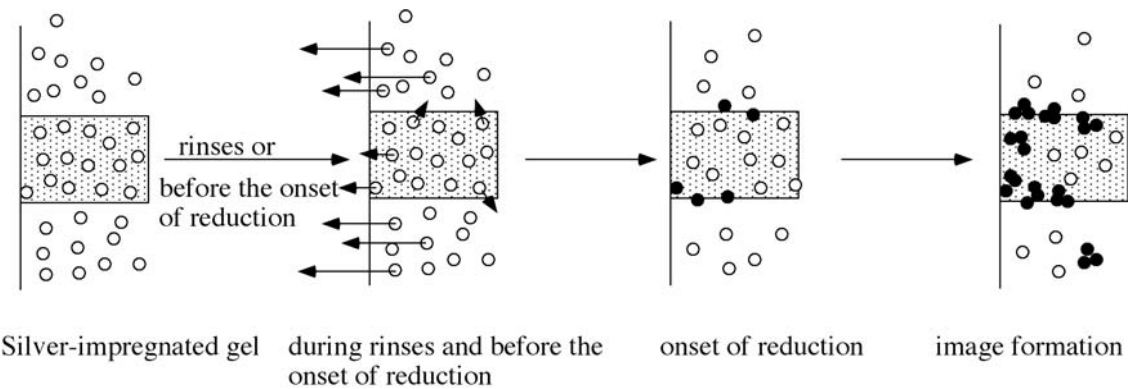

Figure 2

Schematic representation of a positive silver stain. Symbols as in Figure 1. The long arrows show the important silver ion diffusion out of the gel, while the short arrows show the weak diffusion from the proteins due to silver binding.

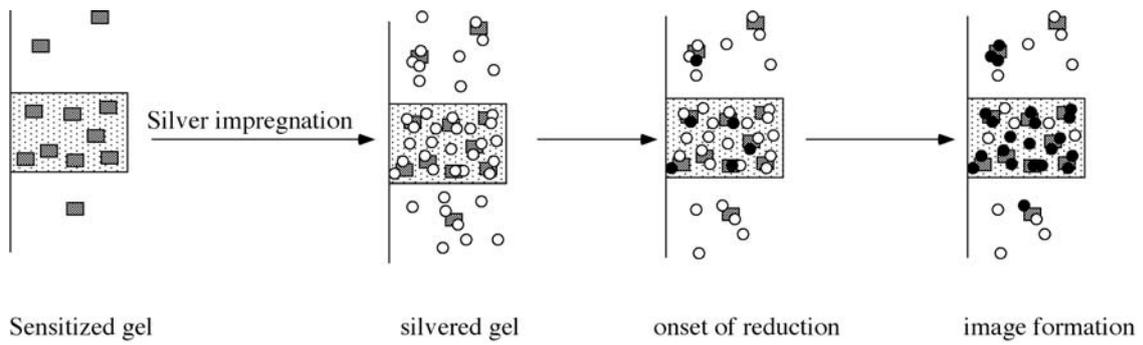

Figure 3

Schematic representation of sensitization by amplification of silver binding. Symbols as in figure 1. the rectangles symbolize the sensitizing agent

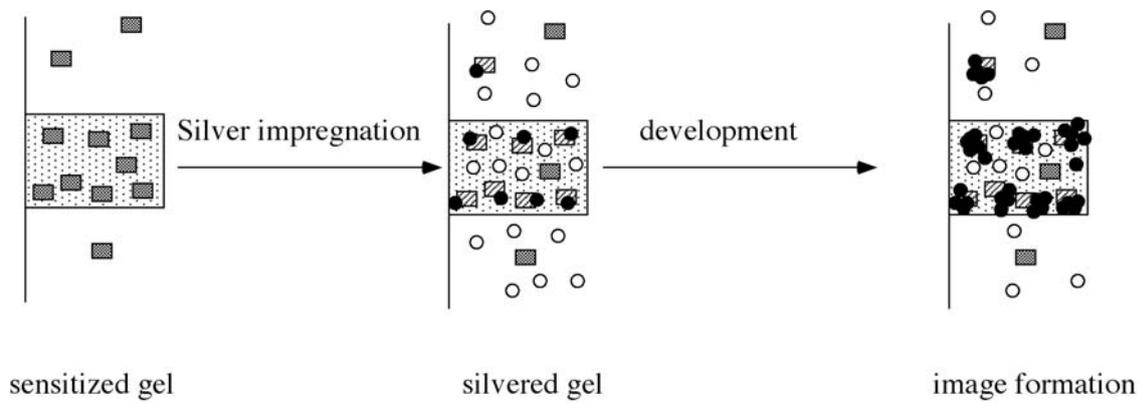

Figure 4

Schematic representation of sensitization by latent image formation. Symbols as in figure 1. the rectangles symbolize the sensitizing agent.

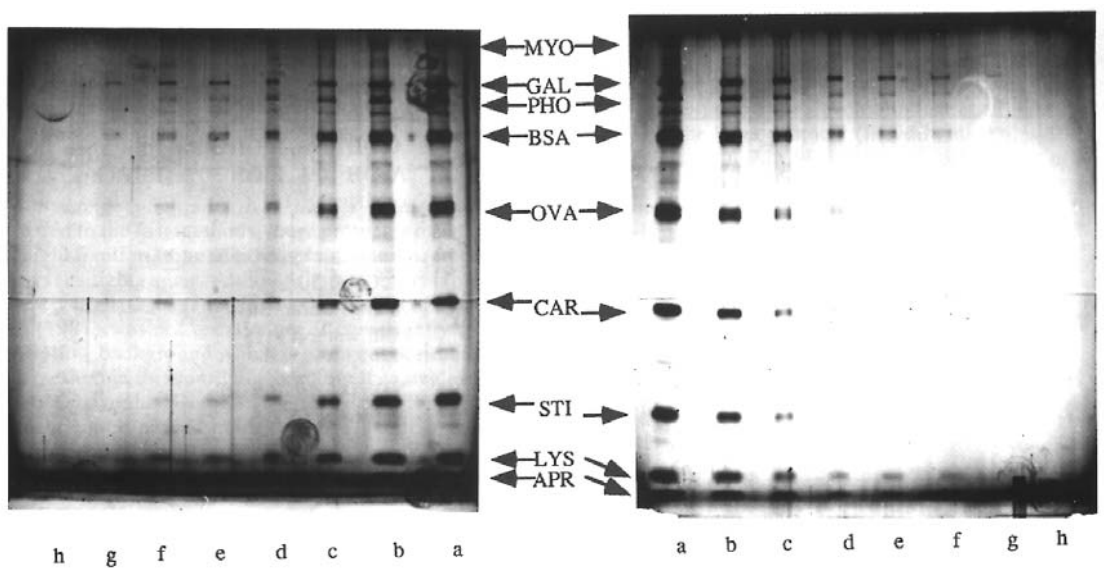

Figure 5: Staining of molecular weight standards using silver nitrate and silver -ammonia stains. Serial dilutions of molecular weight markers (broad molecular weight range from Bio-Rad) were loaded on alternate lanes (1cm wide, giving a mean band area of 10 mm$^2$) on a standard acrylamide-Bis gel or an acrylamide/PDA/thiosulfate gel for silver ammonia staining.. From top to bottom: MYO: myosin(200 kDa); GAL: ß-galactosidase (112 kDa); PHO: phosphorylase A (94 kDa),; BSA: bovine serum albumin (69 kDa); OVA: ovalbumin (43 kDa); CAR: carbonic anhydrase (30 kDa); STI: Soybean trypsin inhibitor (21 kDa); LYS: hen egg white lysozyme (14kDa). APR aprotinin (6.5 kDa)

Protein loads : lanes a: 100 ng per band; lanes b: 50 ng; lanes c: 20 ng; lanes d: 10 ng; lanes e: 5 ng; lanes f: 2 ng; lanes g: 1ng; lanes h: 0.5 ng.

left panel: gel stained with a long silver nitrate method (table 3 col A)

right panel: gel stained with a silver ammonia method (Table 3 col B)

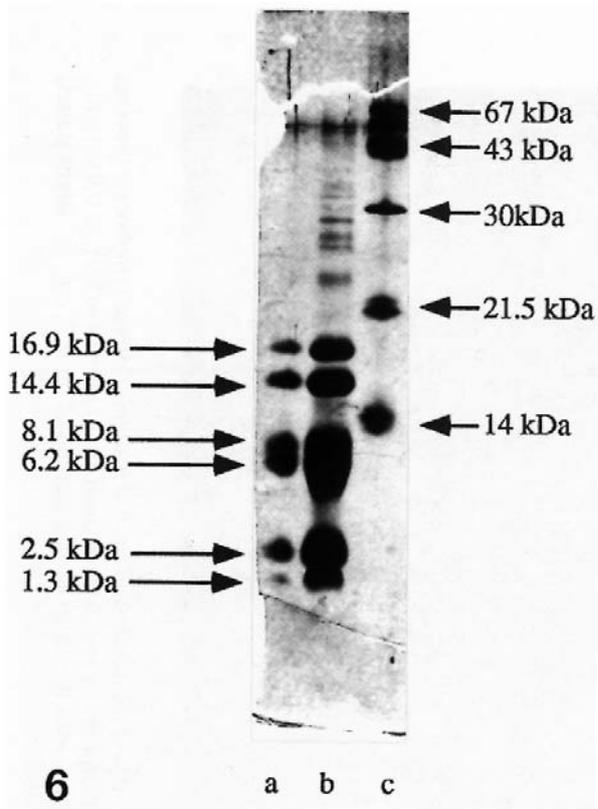

Figure 6: Staining of low molecular mass peptides after separation by a SDS-Tricine gel .

Lanes a and b, Low molecular mass markers (myoglobin peptides from BDH). a: 200 ng total load, b: 400 ng total load

lane c:

Staining as described in Table 2 col C

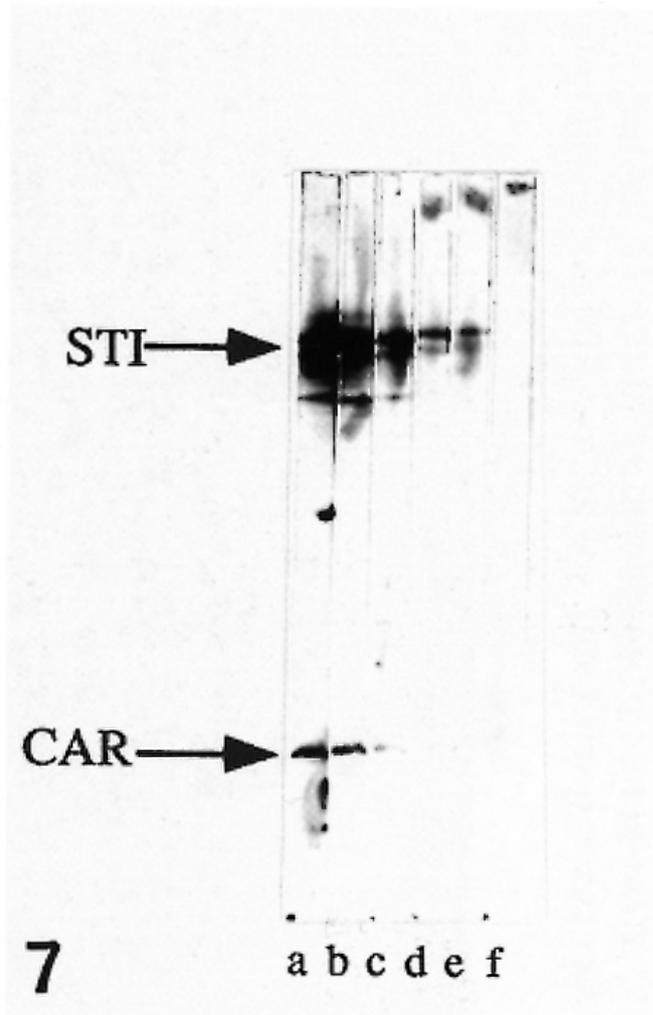

Figure 7: Staining of marker proteins in an IEF gel

STI: soybean trypsin inhibitor (pI 4.9); CAR: carbonic anhydrase (pI 5.9)

a: 1000 ng per marker; b: 500 ng; c: 200 ng; d: 100ng; e: 50 ng; f: 20 ng

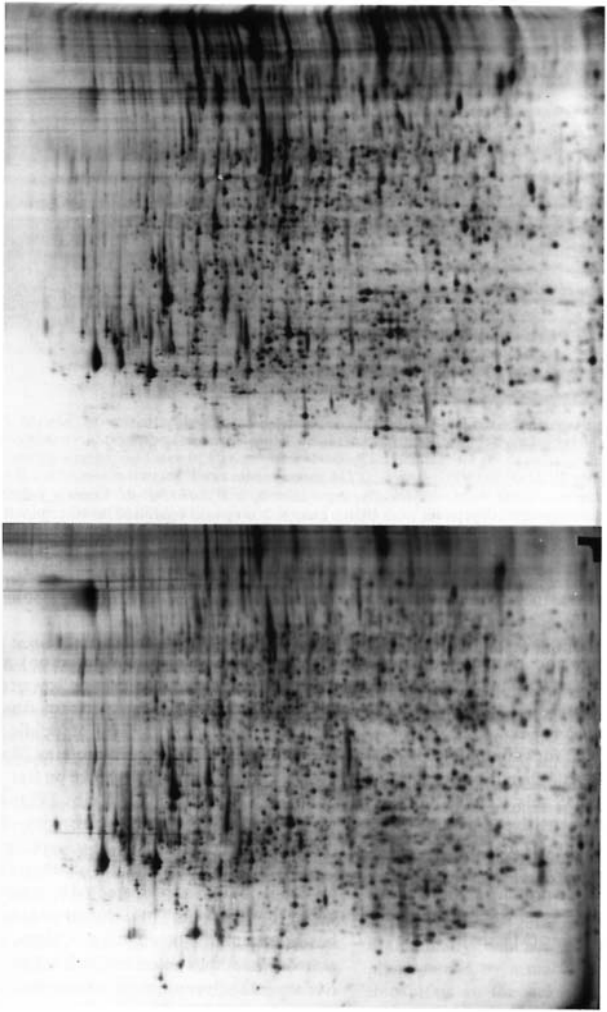

Figure 8: Staining of complex protein samples analysed by two-dimensional electrophoresis. Acidic and neutral proteins.

Mouse lymphocyte proteins (50μg loaded on the IEF gel) were analysed by equilibrium two-dimensional electrophoresis and silver stained. pH gradient 4 to 8 (from left to right), Molecular weight range 250 to 20 kDa (from top to bottom).

Left panel: stained with the silver nitrate method of Table 3 col A; right panel: stained by the silver ammonia method of Table 3, col. B

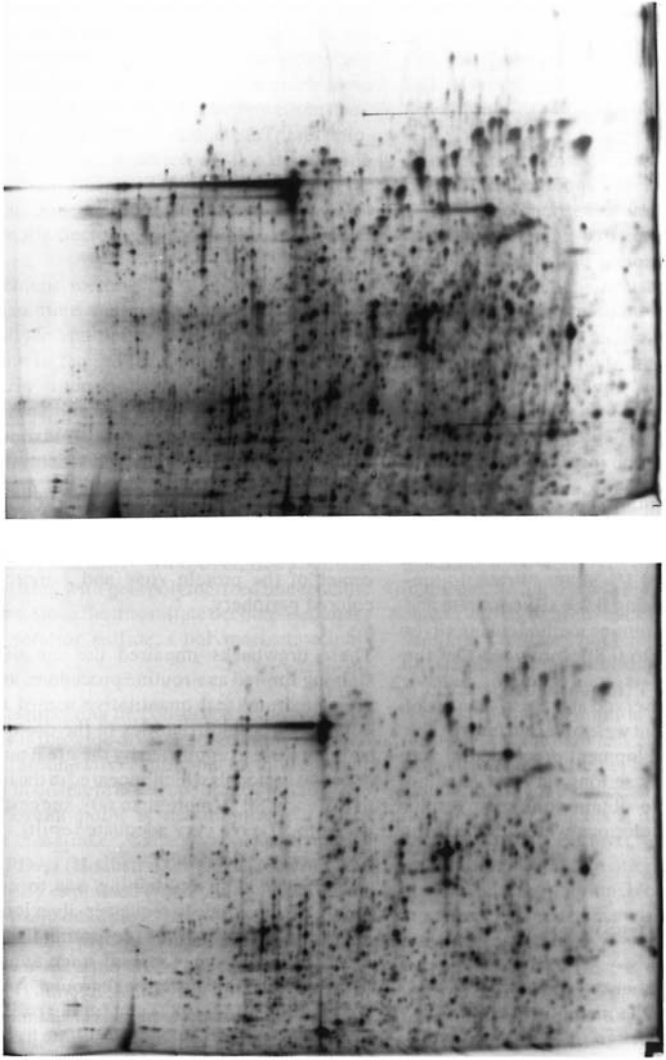

Figure9: Staining of complex protein samples analysed by two-dimensional electrophoresis. Basic proteins.

Mouse lymphocyte proteins (50μg loaded on the IEF gel) were analysed by equilibrium two-dimensional electrophoresis and silver stained. Molecular weight range 250 to 20 kDa left panel: stained by the silver nitrate method of Table 3 col A; right panel: stained by the silver ammonia method of Table 3, col. B

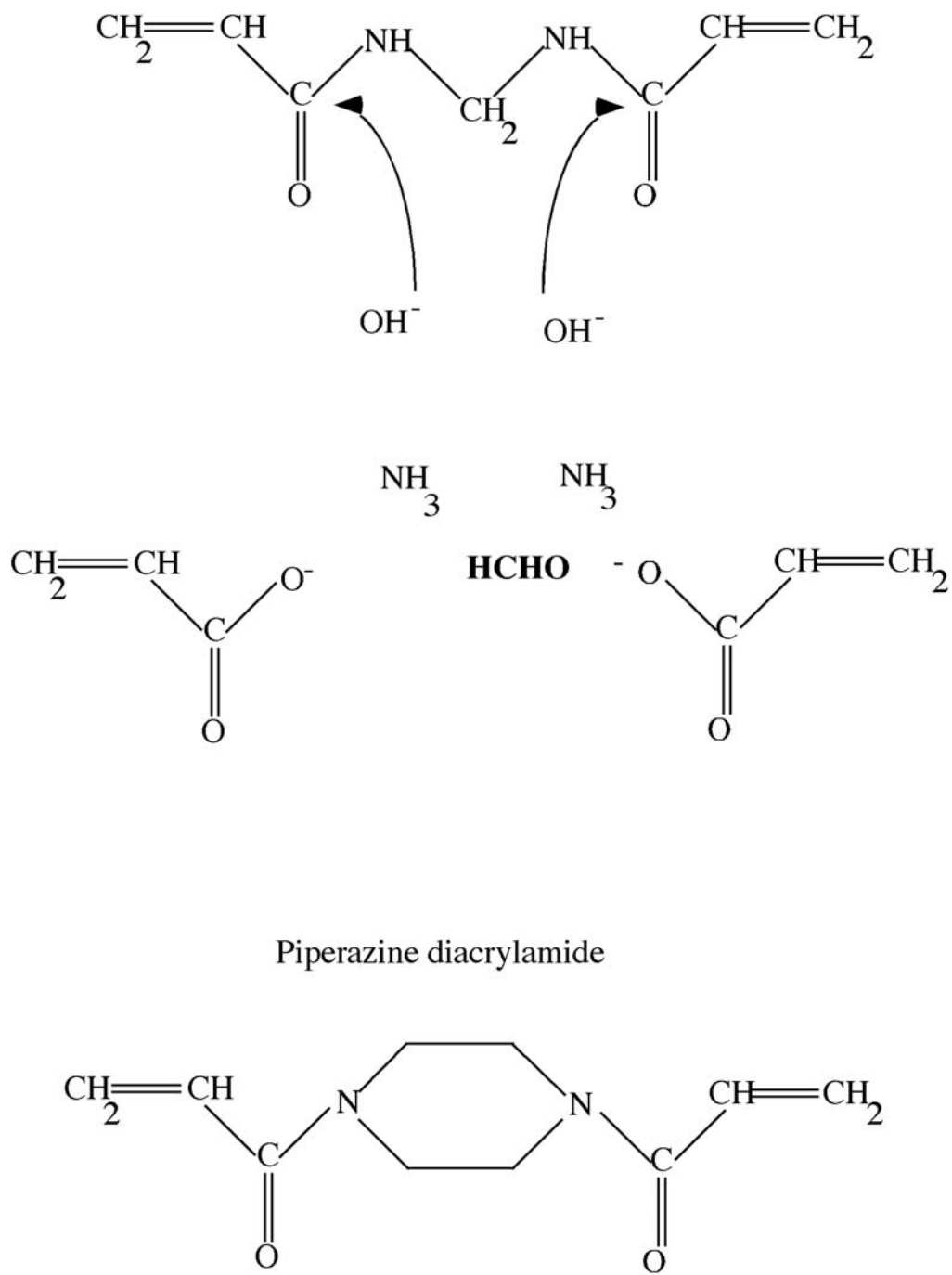

Figure 10:

representation of the alkaline hydrolysis of Bis, showing the formaldehyde production.